\documentclass[pra,twocolumn,floatfix,superscriptaddress]{revtex4-1}
\usepackage{dcolumn,amsmath}
\usepackage{graphicx}
\usepackage{braket}
\usepackage{bm}
\usepackage{amssymb}
\usepackage{hyperref}
\usepackage{multirow}
\usepackage{footnote}
\usepackage{subcaption}
\usepackage{ragged2e}
\usepackage{xcolor}
\usepackage{pdfcomment}
\usepackage[justification=justified,labelfont=bf,singlelinecheck=off]{caption}
\usepackage{threeparttable,booktabs} 
\usepackage{subcaption,siunitx,booktabs}
\usepackage{caption,afterpage,tabularx}
\newcommand{\RUG}{
Van Swinderen Institute for Particle Physics and Gravity,
\\University of Groningen, Nijenborgh 4, 9747AG Groningen, The Netherlands\\
}
\newcommand{\MBU}{
Department of Chemistry, Faculty of Natural Sciences, 
\\Matej Bel University, Tajovsk{\'e}ho 40, 97401 Bansk{\'a} Bystrica, Slovakia\\
}

\newcommand{\TAU}{School of Chemistry, 
Tel Aviv University, 6997801 Tel Aviv, Israel
}

\newcommand{\MUA}{Centre for Theoretical Chemistry and Physics, New Zealand Institute for Advanced Study,
\\Massey University Auckland, Private Bag 102904, 0745 Auckland, New Zealand
}

\newcommand{\CAS}{Centre for Advanced Study (CAS) at the Norwegian Academy of Science and Letters, Drammensveien 78, NO-0271 Oslo, Norway
}

\newcommand{\JGU}{Helmholtz Institute Mainz, 
Johannes Gutenberg University, 55099 Mainz, Germany
}

\newcommand{\UNW}{School of Physics, 
University of New South Wales, Sydney, New South Wales 2052, Australia
}

\begin{document}

\title{The nuclear anapole moment interaction in BaF from relativistic coupled cluster theory}

\author{Yongliang Hao}
\affiliation{\RUG}
\author{Miroslav Ilia\v{s}}
\affiliation{\MBU}
\author{Ephraim Eliav}
\affiliation{\TAU}
\author{Peter~Schwerdtfeger}
\affiliation{\MUA}
\affiliation{\CAS}
\author{Victor V. Flambaum}
\affiliation{\JGU}
\affiliation{\UNW}
\author{Anastasia Borschevsky}
\email{a.borschevsky@rug.nl}
\affiliation{\RUG}


\begin{abstract}

We present high accuracy relativistic coupled cluster calculations of the $P$--odd interaction coefficient $W_A$ describing the nuclear anapole moment effect on the molecular electronic structure. The molecule under study, BaF, is considered a promising candidate for the measurement of the nuclear anapole moment, and the preparation for the experiment is now underway [Altuna{\c{s}} \textit{et al.}, Phys. Rev. Lett. \textbf{120}, 142501 (2018)]. Influence of various computational parameters (size of the basis set, treatment of relativistic effects, and treatment of electron correlation) on the calculated $W_A$ coefficient is investigated and a recommended value of $147.7$ Hz with an estimated uncertainty of $1.5$\% is proposed.

\end{abstract}

\maketitle

\section{Introduction}

The standard model (SM) of particle physics, unifying the electromagnetic, weak, and strong forces by which the fundamental particles interact, has proven to be valid within unprecedented accuracy. However, the SM as we know is incomplete and many open questions remain that lie beyond its current formulation \cite{Pat16}; among the most important are the origin of dark matter and dark energy, neutrino mass and oscillations, matter-antimatter asymmetry and the unification with the gravitational force. These open questions motivate both formulations of new theories beyond the SM and experimental searches for new physical phenomena. 

One prominent category of such experiments is the search for violation of parity ($P$) and time ($T$) reversal symmetries in atoms and molecules \cite{SafBudDem17}. In particular, investigation of nuclear spin dependent parity violating (NSD-PV) effects, which are the main focus of this work, can be used to test low-energy quantum chromodynamics and parity nonconservation in nuclei \cite{GinFla04}.

The NSD-PV term of the electronic Hamiltonian for a specific nucleus can be written as \cite{SafBudDem17}, 
\begin{equation}
{H_{{\rm{NSD}}}} = \,\frac{{{G_\mathrm{F}}}}{{\sqrt 2 }I} \sum_i \left( {{\kappa_{{\rm{A}}}} + {\kappa _{{\rm{ax}}}} + {\kappa_{{\rm{hfs}}}}} \right)\, \left(\bm{\alpha}_i \cdot \,{\bm{I}} \right) \,{\rho} \left( \bm{r}_i \right),
\label{HNSD}
\end{equation}
where $G_{F}\approx 1.435850(1) \times 10^{-62}$ J$\cdot$m$^3 \approx 2.222516(1)\times10^{-14}$ $E_h \cdot a_B^3$ is the Fermi
coupling constant \cite{Mohr2016}, $E_h$ is the Hartree energy, $a_B$ is the Bohr radius, $\bm{\alpha}$ the Dirac matrices in the standard representation, $\bm{I}$ the nuclear spin, $\bm{r}_i$ the electronic coordinates for electron $i$, and $\rho(\bm{r}_i)$ the (normalized) nuclear density distribution. This contribution is only present for nuclei with $I \neq 0$, and for open-shell atoms or molecules because of Kramers symmetry. The three dimensionless nuclear $\kappa$ parameters are associated with the different sources of the NSD-PV effects. The first term, $\kappa_A$, comes from the nuclear anapole moment interaction and will be discussed in more detail below. The second term $\kappa_\textrm{ax}$ arises from the electroweak neutral coupling between the electron vector and nucleon axial-vector currents ($\mathbf{V}_e\mathbf{A}_N$) \citep{NovSusFla77-1}; theoretical prediction of $\kappa_\textrm{ax}$ within the nuclear shell model can be found in Ref.~\cite{FlaKhr80-1}.  
The third contribution $\kappa_\textrm{hfs}$ originates in the nuclear-spin-independent weak interaction combined with the hyperfine interaction \cite{FlaKhr85-1}. The coefficients $\kappa_\textrm{hfs}$ were derived using different models, for example in Refs. \cite{FlaKhr85_2,BouPik91,JohSafSaf03}. 

The anapole moment was first predicted by Zel'dovich \cite{Zel58} in 1958. It  appears in the
second-order multipole expansion of the magnetic vector-potential simultaneously  
with the $P$-- and $T$-- violating  magnetic quadrupole moment
\cite{SusFlaKhr84}. In a simple  valence  nucleon model $\kappa_\textrm{A}$ has the following form \cite{FlaKhrSus84},
\begin{eqnarray}
\kappa_\textrm{A}= 1.15\times 10^{-3} \left(\frac{\mathcal{K}}{I+1} \right) A^{\frac{2}{3}} \mu_i g_i.
\label{eqanapole}
\end{eqnarray}
Here, $\mathcal{K}=(-1)^{I+\frac{1}{2}-l}(I+1/2)$, $l$ is the orbital
angular momentum of the external unpaired nucleon $i=n,p$; $\mu_p= +2.8$, $\mu_n= -1.9$, and $A$ is the atomic mass number. Theoretical estimates give the dimensionless strength constant for nucleon-nucleus weak potential $|g_p| \approx 4.6$ for a proton \cite{FlaKhrSus84}, and $|g_n|\sim 1$ for a neutron \cite{FlaMur97}. Due to the $A^{2/3}$ scaling of this effect, the nuclear anapole moment provides the dominant NSD-PV contribution for systems containing heavy nuclei \cite{GinFla04}. The determination of nuclear anapole effects can contribute to the fundamental understanding of parity violation in the hadronic sector \cite{HaxWie01,GinFla04}.

To date, only one observation of a non-zero nuclear anapole moment was achieved using a Stark-PV interference technique in an experiment on the $^{133}$Cs atom \cite{WooBenCho97}, where the main source of the anapole moment was due to the unpaired proton. The value of $\kappa_{\rm A}$ for $^{133}$Cs was determined as $\kappa_A$=364(62)$\times$10$^{-3}$ \cite{FlaMur97}.
Further measurements on Cs and other alkali atoms using the ground state hyperfine splitting have been recently proposed \cite{ChoEli16}.
Complementary measurements are also being performed on atoms with unpaired neutrons, such as $^{171}$Yb \cite{AntFabBou17} and $^{212}$Fr \cite{AubBehCol13}.

It was shown early on \cite{Lab78,SusFla78,FlaKhr85_2} that NSD parity violating effects are strongly enhanced in diatomic molecules with  $^2\Sigma_{1/2}$ and
$^2\Pi_{1/2}$ electronic states due to the mixing of close rotational
states of opposite parity. Thus, these systems provide a different, advantageous route for the search for these phenomena. An experiment to measure NSD-PV effects using the Stark-PV interference
technique in polar diatomic molecules was proposed in 2008 by \citet{DeMCahMur08}. In this approach, the opposite parity rotational or hyperfine levels of ground state molecules are tuned to near-degeneracy by a magnetic field, causing dramatic amplification of the parity violating effects \cite{KozLabMit91}. Highly sensitive measurements of this type using the $^{138}$BaF molecule were demonstrated recently \cite{AltAmmCah18-1,AltAmmCah18-2}. Another experiment based on optical rotation measurements in $^{199}$HgH was also proposed \cite{GedSkrBor18}.

In diatomic molecules with non-zero nuclear spin and $^{2}\Sigma _{1/2}$ and $^{2}\Pi _{1/2}$ electronic states the nuclear anapole moment interaction can be rewritten in a slightly simplified form,

\begin{eqnarray}
H_{\text{A}}=\kappa _{\text{A}}\ \frac{G_{\mathrm{F}}}{\sqrt{2}}\sum_{i}\rho
(\bm{r}_{i}){\alpha _{+}}
\end{eqnarray}

\noindent with
\begin{eqnarray}
\alpha _{+}=\alpha _{x}+i\alpha _{y}=\left(
\begin{array}{cc}
{0} & \sigma _{x} \\
\sigma _{x} & {0} \\
\end{array}%
\right) +i\left(
\begin{array}{cc}
{0} & \sigma _{y} \\
\sigma _{y} & {0} \\
\end{array}%
\right).
\end{eqnarray}

\noindent Here, $\sigma_x$ and $\sigma_y$ are the Pauli matrices. The $^{2}\Sigma_{1/2}$ and $^{2}\Pi _{1/2}$ open-shell electronic states are twofold degenerate, corresponding to the two possible projections of electronic angular momentum along $\bm{n}$, i.e., $\Ket\Omega =\Ket{\pm \frac{1}{2}}$, where $\bm{n}$ is the unit vector directed along the molecular axis from the heavier to the lighter nucleus. The interaction $H_{\text{A}}$ removes the degeneracy and mixes $\Ket\Omega $ states with
different signs (parities). 

The $P$--odd interaction coefficient $W_{\text{A}}$ is usually explored for the expression of the strength of coupling of
the two different parity states. This coefficient depends on the electronic structure of the molecule and is defined for a given electronic state; it can be derived from the expression for $H_{\text{A}}$ as the transition element between the two different $\Ket\Omega $ states \cite{KudPetSkr14}, 
\begin{eqnarray}
W_{\text{A}}=\frac{G_{\mathrm{F}}}{\sqrt{2}}\bra {+\tfrac{1}{2}}\sum_{i}\rho
(\bm{r}_{i}){\alpha _{+}}\ket{-\tfrac{1}{2} }.
\end{eqnarray}
Note that the matrix elements calculated between the same $\Ket\Omega $
states are zero. Thus, the coefficient $W_{\text{A}}$ defines the amplitude
of the expectation value of $H_{\text{A}}$ in the mixed-parity state.
Knowledge of $W_\textrm{A}$ is required for extracting the nuclear anapole moment from experiment. It can not be measured directly, and  has to be provided by theory. Needless to say the accuracy and reliability of the calculated $W_\textrm{A}$ coefficients is important for the meaningful interpretation of any measurement, and it is thus most desirable to employ state-of-the-art relativistic quantum theoretical methods for such calculations. 

Here we perform relativistic coupled cluster calculations to obtain the $W_\textrm{A}$ coefficient for BaF within the framework of a finite-field approach. We investigate the sensitivity of $W_\textrm{A}$ to various computational parameters allowing us to estimate the uncertainty of our result, and finally propose a recommended value for interpretations of future experiments on this molecule.

To the best of our knowledge, only two previous studies used coupled cluster theory for the calculation of the $W_\textrm{A}$ coefficient. Relativistic two-component Fock-space coupled cluster theory was used to calculate $W_\textrm{A}$ and other $P$- and $T,P$--odd parameters for RaF (the authors estimated the uncertainty of the results as 10\%) \cite{KudPetSkr14}. More recently, relativistic Fock-space coupled cluster method was used to calculate the $W_\textrm{A}$ coefficients of the $^2 \Sigma _{1/2}$ and the $^2 \Pi _{1/2}$ electronic states of HgH \cite{GedSkrBor18}.

The majority of earlier investigations of NSD-PV effects in diatomic molecules such as BaF have relied on more approximate approaches such as semiempirical methods \cite{KozLab95, DeMCahMur08}, where the $W_\textrm{A}$ parameters were estimated using experimental spectroscopic data. Kozlov \textit{et al.} performed relativistic effective core potential (RECP) calculations in the framework of a self-consistent-field (SCF) approach estimating core polarization effects by an effective operator (EO) \cite{KozTitMos97}.
Nayak and Das \cite{NayDas09} carried out Dirac-Hartree-Fock (DHF) calculation within restricted active space configuration interaction (RASCI). Isaev and Berger \cite{IsaBer12} used a quasirelativistic two-component zeroth-order regular approximation (ZORA) combined with Hartree-Fock (HF) and density functional theory (DFT), and scaled the results using a semiempirical model described in Ref. \cite{Koz85}. We have previously carried out both DHF and DFT calculations of this property for BaF and many other diatomic molecules \cite{BorIliDzu12-1,BorIliDzu13}. In that work the average of the DHF and the DFT results scaled by the effect of core-polarization (CP) obtained from atomic calculations was taken as the recommended value; these results are designated here as DHF/DFT+CP. 

\section{Method and computational details}

The calculations were carried out using the adapted version of the DIRAC program package \cite{DIRAC15} in the framework of the Dirac-Coulomb Hamiltonian,

\begin{eqnarray}
H_{0} = \sum\limits_{i}[c\boldsymbol\alpha_i \cdot \bm{p}_{i} + \beta_i m c^2 + V(r_i)] + \sum\limits_{i<j} \frac{1}{r_{ij}},
\end{eqnarray}
where $\boldsymbol\alpha_i$ and $\beta$ are the Dirac matrices in standard representation. The Coulomb potential $V(r_i)$ takes into account the finite size of the nuclei, modelled by Gaussian charge distributions \cite{VisDya97}.

The $P-$odd interaction constant is a property of a given nucleus within the molecular environment and for a diatomic molecule we have two $W_\textrm{A}$ values. In this work, however, we entirely focus on the $W_A$ parameter at the metal nucleus relevant for future experiments. 

In order to perform coupled cluster calculations for the $W_\textrm{A}$ parameter,
we employ a finite-field  approach (FF) \cite{Mon77,Thyssen2000}. Within this scheme,
the entire Hamiltonian of the system $H$ is regarded to be a function of some
perturbation parameter $\lambda$,

\begin{eqnarray}
H(\lambda )=H_{0}+\lambda \frac{G_{\mathrm{F}}}{\sqrt{2}}\sum_{i}\rho (\bm{r}%
_{i}){\alpha _{+}}.
\end{eqnarray}

\noindent The nuclear density $\rho(\bm{r}_i)$ is of Gaussian shape, which is suitable for the fully relativistic framework of the present work. For small values of $\lambda$, the total energy can be expanded in Taylor series around $%
\lambda=0$,

\begin{eqnarray}
E(\lambda)=E(0)+\lambda\frac{dE(\lambda)}{d\lambda}\bigg|_{\lambda=0}+...
\end{eqnarray}

The calculations are performed at various perturbation strengths $\lambda$. If these are chosen to be small enough to remain in the linear regime, the higher order terms can be ignored and $W_\textrm{A}$ can be obtained numerically, according to the Hellmann-Feynman theorem, from the first derivative of the energy with respect to $\lambda$:
\begin{eqnarray}
W_{A}=\frac{dE(\lambda )}{d\lambda }\bigg|_{\lambda =0}.
\end{eqnarray}
The perturbation strength needs to be sufficiently large such that the change in total energy is not lost in the precision of the calculations. We have tested the linearity of the above expression with different perturbation strengths applied, i.e. $\lambda = 10^{-6}, 10^{-7}, 10^{-8}$ and $10^{-9}$. Based on our results, we found that minimal error in linear fit is obtained for perturbation strengths of the order of $\sim 10^{-8}$. Furthermore, the energy convergence requirement of the coupled cluster iterations had to be set to $10^{-12}$ a.u. 

We have used and compared two variants of relativistic coupled cluster theory: the standard single-reference coupled cluster method with single, double, and perturbative triple contributions, CCSD(T) \cite{VisLeeDya96}, and the multireference Fock-space coupled cluster approach (FSCC) \cite{VisEliKal01}. Within the framework of the valence-universal FSCC approach an effective Hamiltonian is defined and calculated in a low-dimensional model (or $P$) space, constructed from zero-order wave functions (Slater determinants), with eigenvalues approximating some desirable eigenvalues of the Hamiltonian. According to Lindgren's formulation of the open-shell CC method \cite{Lin89}, the effective Hamiltonian has the form

\begin{eqnarray}
H_{Eff}=PH \Omega P,\;  \Omega=\exp{S},
\end{eqnarray}
where $\Omega$  is the normal-ordered wave operator and the excitation operator $S$ is defined with respect to a closed-shell reference determinant (vacuum state) and partitioned according to the number of valence holes ($m$) and valence particles ($n$) with respect to this reference:

\begin{eqnarray}
S= \sum_{m \geqslant 0} \sum_{n \geqslant 0} \Big( \sum_{l \geqslant m+n} S_l^{(m,n)} \Big).
\end{eqnarray}
Here $l$ is the number of excited electrons.

BaF has a single valence electron occupying the $\sigma$ orbital and thus two different computational schemes are appropriate for this system. In the first scheme, designated FSCC(0,1), we start with BaF$^+$. After solving the relativistic Dirac-Fock equations and correlating this closed shell reference state, an electron is added to reach the neutral state. At each stage the coupled cluster equations are solved to obtain the correlated ground and excited state energies. The extra electron can be added to the lowest $\sigma$ orbital, or allowed to also occupy the higher states, thus yielding a number of energy levels and also improving the description of the ground state energy and properties. We have tested the influence of the size of the model space $P$ on the calculated $W_\textrm{A}$ parameters. Within the second computational scheme (FSCC(1,0)) the calculation begins from the closed shell negative ion 
BaF$^-$, and an electron is removed to obtain the neutral system. In principle, the two schemes should give very similar results for the ground state, the main difference stemming from the different closed shell reference states yielding different Hartree-Fock orbitals (i.e. relaxation effects). 
In addition to the coupled cluster results, we also report $W_\textrm{A}$ values from second-order M\o{}ller-Plesset perturbation theory, MP2 \cite{MolPle34}.

In order to further investigate the effects of electron correlation we performed open-shell
single determinant average-of-configuration DHF \cite{Thy01} and relativistic DFT \cite{SauHel02} calculations for $W_\textrm{A}$ by evaluating the matrix elements of the  $\boldsymbol\alpha \rho(\mathbf{r}_i)$ operator in the molecular spinor basis. 
To test the performance of various functionals for this property the DFT calculations were carried out with the Perdew-Burke-Ernzerhof (PBE) functional \cite{PerBurErn96,BurPerErn97}, the Slater local exchange (SVWN5) functional \cite{VosWilNus80}, the Becke-Lee-Yang-Parr hybrid functional (B3LYP) \cite{LeeYanPar88,Bec93,Bec99} and its Coulomb-attenuated version (CAMB3LYP*), adapted to accurately describe PV energy shifts in heavy atomic systems obtained from coupled cluster theory \cite{YanTew04,ThiRauSch10}. 

Standard Dyall's basis sets of varying size \cite{Dya09,Dya16} were employed to investigate the basis set effects on the calculated $W_\textrm{A}$ values. To further improve our results we augmented the basis sets of the two atoms by additional large (tight) and small (diffuse) exponent functions (see below for details). Further investigated computational parameters were the active space in the electron correlation procedure, i.e. the number of correlated electrons and the chosen virtual energy cut-off. In addition we include the Gaunt term in our calculations \cite{Gau29} as part of the Breit interaction, which corrects the 2-electron part of the Dirac--Coulomb Hamiltonian up to order $(Z \alpha)^2$ \cite{Bre29}. The Gaunt interaction is included self-consistently at the DHF step. Alongside the detailed investigations of BaF, we also perform calculations for the $W_\textrm{A}$ parameters of its lighter homologues BeF, MgF, CaF, and SrF in order to examine the dependence of $W_\textrm{A}$ on the nuclear charge of the Group 2 atom. 
The positions of the atoms were chosen according to the molecular experimental equilibrium bond lengths (1.361 $\mbox{\AA}$ for BeF, 1.750 $\mbox{\AA}$ for MgF, 1.967 $\mbox{\AA}$ for CaF \cite{NIST}, 2.076 $\mbox{\AA}$ for SrF \cite{BarBea67}, and 2.159 $\mbox{\AA}$ for BaF  \cite{RyzTor80}).

\section{Results and discussion}

The first important step in our investigation was a detailed study of the influence of the basis set size on the $W_\textrm{A}$ parameters; we also use this study to determine the best basis set which is still affordable computationally. These tests were performed within the DHF, CCSD, and CCSD(T) framework. In the coupled cluster calculations, 35 electrons were correlated and virtual orbitals with energies above 30.0 a.u. were excluded. We used the standard Dyall's relativistic basis sets of double-, triple-, and quadruple-zeta quality \cite{Dya09,Dya16}. To check the influence of diffuse functions, we have augmented the dyall.v4z basis with a single diffuse function for each  symmetry (s-aug-dyall.v4z) and with two diffuse functions (d-aug-dyall.v4z). While diffuse functions are usually more important for chemical properties, a good description of the electronic wave function in the nuclear region is essential for obtaining reliable results
for parity-violating effects \cite{LaeSch99}. In particular, it was demonstrated in our earlier work \cite{BorIliDzu12-1,BorIliDzu12-2,BorIliDzu13} that tight $s$ and $p$ functions have a considerable influence on the $W_\textrm{A}$ parameter at the DHF level, especially for the lighter elements. Thus, we also tested the effect of adding different types of tight functions to the basis sets (designated as t$s$ for high exponent $s$ function, t$p$ for high exponent $p$ and etc.). The augmentations (both with the diffuse and the tight functions) were carried out separately for each of the atoms, and the results are summarised in Table ~\ref{TabI}. 

Going from double- to triple-zeta quality basis set increases the calculated $W_\textrm{A}$ value by $\sim$10\%; moving to quadruple zeta quality leads to a further increase of less than a single percent on the coupled cluster level. The correlation part of the $W_A$ coefficient does not scale smoothly with the size of the basis set, and hence we did not perform extrapolation to the complete basis set limit.  Adding diffuse functions for barium has negligible effects on the results, while augmenting the basis of fluorine reduces the $W_\textrm{A}$ value by 0.5\% as the fluorine orbitals extend into the the domain of the Ba atom. Out of the large exponent functions, only the tight $f$-type function has a discernible influence on the calculated $W_\textrm{A}$, raising its value by $\sim$1\% on CCSD and CCSD(T) level (but having no impact on the DHF results). Adding a second tight $f$ orbital leaves the calculated $W_\textrm{A}$ almost unchanged. 
We thus assume that the results are converged (close to the basis set limit) and perform the rest of our calculations using the optimized dyall.v4z+t$f$ basis set.

\begin{table}[t] 
\caption{Basis set dependence of the calculated $W_\textrm{A}$ coefficient (Hz)  of BaF.}\label{Topt}
\begin{ruledtabular}
\begin{tabular}{llrcc}
Basis (Ba)&Basis (F)&DHF&CCSD&CCSD(T)\\\hline
v2z            &v2z            & 99.12  &  131.04  &  129.22  \\
v3z            &v3z            &110.70  &  143.51  &  141.42  \\
v4z            &v4z            &112.27  &  144.02  &  141.84  \\\hline
\multicolumn{5}{c}{Diffuse functions}    \\
s-aug-v4z      &dyall.v4z      &112.25  &  143.95  &  141.76  \\
d-aug-v4z      &dyall.v4z      &112.25  &  143.95  &  141.76  \\
dyall.v4z      &s-aug-v4z      &112.09  &  143.36  &  141.16  \\
dyall.v4z      &d-aug-v4z      &112.04  &  143.27  &  141.05  \\\hline
\multicolumn{5}{c}{Tight functions}    \\
v4z+t$s$         &v4z          &112.30  &  144.08  &  141.89  \\
v4z+t$p$         &v4z          &112.39  &  144.15  &  141.97  \\
v4z+t$d$         &v4z          &112.27  &  144.00  &  141.82  \\
v4z+t$f$         &v4z          &112.32  &  145.17  &  143.02  \\
v4z+2t$f$         &v4z         &112.33  &  145.28  &  143.08  \\
v4z+t$g$         &v4z          &112.27  &  144.42  &  142.31  \\
v4z            &v4z+t$s$       &112.27  &  144.02  &  141.84  \\
v4z            &v4z+t$p$       &112.27  &  144.02  &  141.84  \\
v4z            &v4z+t$d$       &112.27  &  144.00  &  141.82  \\
v4z            &v4z+t$f$       &112.27  &  144.00  &  141.82  \\
\end{tabular}
\end{ruledtabular}
\label{TabI}
\end{table}

\begin{figure}[t]
\centering
\includegraphics[scale=0.75,width=0.99\linewidth]{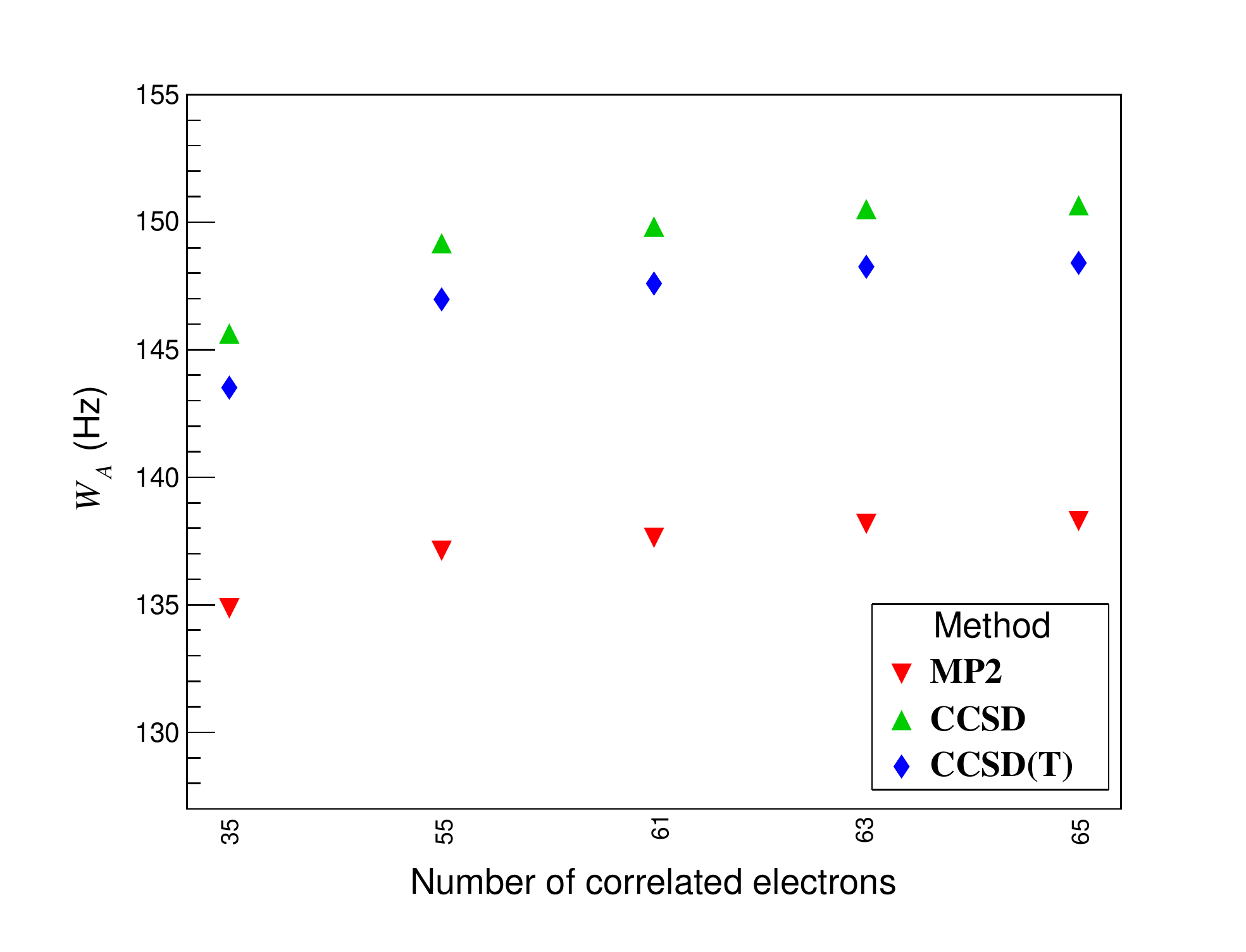}
\caption{Calculated $W_\textrm{A}$ coefficients of BaF using different number of correlated electrons.}
\label{Fcor}
\end{figure}

Next we explored the effect of the number of correlated electrons and the size of the virtual space on $W_\textrm{A}$. 
In the first set of calculations we keep the energy cut-off for the virtual space at a rather high value (500 a.u.) and vary the number of correlated electrons. Figure ～\ref{Fcor} ~presents the calculated MP2, CCSD, and CCSD(T) $W_\textrm{A}$ values. Overall, the difference between including 35 electrons in the calculation (corresponding to the commonly used cut-off of -20.0 a.u. in the space of the occupied orbitals) and correlating all 65 electrons is  $\sim$3\% for the three methods. We note that the major part of this difference does not come from the 1$s$ orbital alone, despite its proximity to the nucleus, but rather comes from all the core shells. In order to achieve $<1\%$ accuracy, all the electrons should be included in the electron correlation procedure.

\begin{figure}[t]
\centering
\includegraphics[scale=0.75,width=0.99\linewidth]{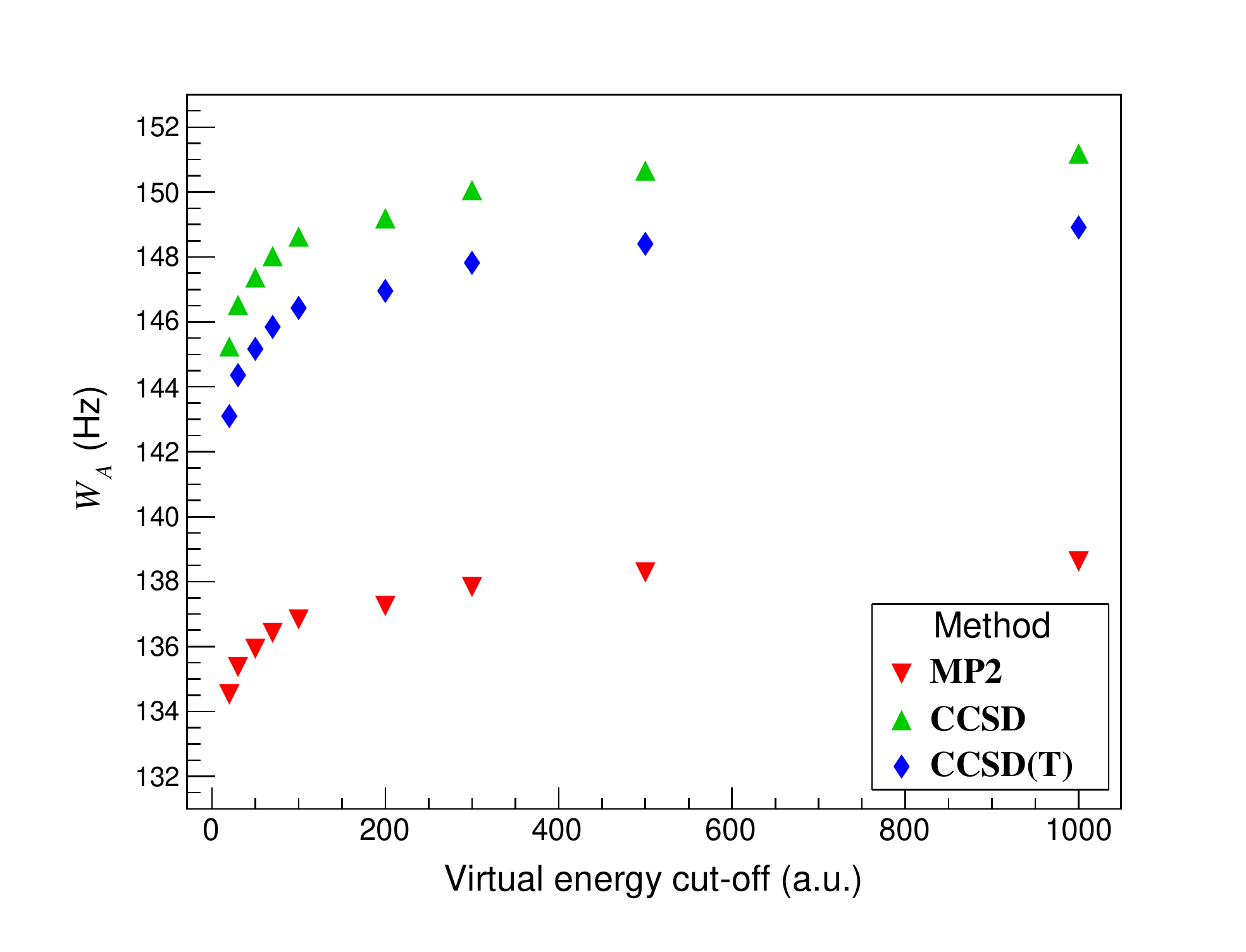}
\caption{Calculated $W_\textrm{A}$ coefficients of BaF using different sizes of the virtual space.}
\label{Fvir}
\end{figure}

In the next step we perform calculations where all electrons are correlated, and vary the energy cut-off in the virtual space (Figure ~\ref{Fvir}). Unlike many other atomic or molecular properties, the calculated $W_\textrm{A}$ value does not saturate at the energy cut-off of about 30 a.u., but continues to increase. The difference in the value corresponding to cut-off of 500 a.u. compared to 30 a.u. is $\sim$3\%, and the $W_\textrm{A}$ value continues to increase further beyond this point, albeit at a much lower rate ($W_\textrm{A}=148.91$ Hz for cut-off of 1000 a.u. vs. 148.40 Hz for 500 a.u.). The importance of inclusion of high lying virtual orbitals for the correlation of the core electrons was also observed by Skripnikov et al. \cite{SkrMaiMos17} for the scalar-pseudoscalar interaction constant $R_s$ in the francium atom.
We selected a final cut-off of 500 a.u. for the following calculations, which is a compromise between optimal accuracy and computational feasibility.

Table~\ref{Tapr} contains the $W_\textrm{A}$ constant of BaF (for the Ba atom), calculated at the DHF, MP2, and DFT levels of theory, using different functionals, and within various coupled cluster schemes. These results were obtained using the optimised basis set (dyall.v4z augmented by a single tight $f$ function) and in the MP2 and CC calculations all electrons were correlated with the energy cut-off for the virtual space set to 500 a.u.

Electron correlation clearly plays an important role for this property, and the CCSD results are 25\% higher than the corresponding DHF values. MP2, however, performs remarkably well and captures the majority of electron correlation, differing from the CCSD value by only 8\%.

DFT results tend to be very close to the DHF value; in particular, the result obtained with the CAMB3LYP* functional,  which is generally expected to perform well for parity-violating properties \cite{ThiRauSch10}, is almost identical to the DHF result, which is somewhat disappointing. Note, similarly disappointing results are obtained for electric field gradients of molecules containing transition metals \cite{Bast-2003}, and arguments about the failure of DFT can be found in Ref. \cite{Schwerdtfeger:1999fk}. As we expect that most density functionals lie in-between the DHF and the local density approximation (SVWN5) results, here this deficiency cannot be so easily fixed by adjusting the Hartree-Fock contribution in the hybrid functional as was done for electric field gradients \cite{Bast-2003}.

Moving to the coupled cluster results, the triple excitations contribute very little and lower the $W_\textrm{A}$ value by $\sim$1.5 \% only. We thus expect that the higher-order excitations in the CC procedure will not play an important role for this property.   

As already mentioned, we have tested two variants of FSCC. In the first one, FSCC(0,1), the calculation starts from BaF$^+$, and an electron is added into the virtual orbitals. Here we test two sizes of the model space: the minimal one, designated Model space I, where the additional electron is allowed to occupy only the lowest $\sigma$ orbital (yielding the ground $X^2 \Sigma_{1/2}$ state), and Model space II, which contains 2 $\sigma$, 2$\pi$, and 2$\delta$ orbitals. 
 
The second FSCC scheme, FSCC(1,0), starts with BaF$^-$ as reference state, and an electron is removed to reach the neutral system. Usually, one expects the FSCC results to be situated between the CCSD and the CCSD(T) values (for recent reviews of relativistic FSCC approach, see Ref.~\cite{EliBorKal17}). The sector (0,1) results are extremely close to the CCSD(T) values; superior performance of FSCC in particle sectors compared to single reference CCSD has been observed in the past \cite{EliBorKal17}. It should be noted that in this case increasing the size of the model space has negligible influence on the results. The sector (1,0) values are slightly higher than the CCSD ones, rather than lower as one would expect. This is probably due to the fact that the (1,0) calculation starts from a negative closed shell reference state, and the basis which was optimised for the neutral system does not provide sufficient description of the more diffuse orbitals in BaF$^-$.

The Gaunt interaction lowers the $W_\textrm{A}$ (on DHF level) by 0.7 Hz. We add the Gaunt contribution from the DHF calculation to the CCSD(T) result to provide the final (recommended) value for the $W_\textrm{A}$ constant of BaF. This value is designated as CCSD(T)+Gaunt in Table~\ref{Tab2}.  

In order to put an error bar on this value, we need to examine the remaining sources of uncertainty within our computational approach. These include basis set deficiencies, the unaccounted full triple and higher order contributions in the coupled cluster procedure, the choice of the virtual space cut-off, and neglect of the full Breit and higher order QED effects. From the investigation of the basis set effects (Table ~\ref{TabI}) we see that the contribution from the diffuse functions on the F atom, which we neglect here, is around $-0.6$ Hz, while the effect of tight functions beyond tight $f$ (mostly stemming from the tight $g$) is $+0.5$ Hz. These two effects cancel out, but we take a conservative estimate of basis set uncertainty of about 0.5 Hz (on the order of magnitude of these effects), to account for any further shortfalls of the basis set. The difference in the values of $W_A$ calculated with a virtual space cut-off of $500$ a.u. and $1000$ a.u. is about $0.5$ Hz and it seems that saturation is reached; we thus take $1$ Hz as the corresponding uncertainty. The contribution of perturbative triple excitations in the calculation via the CCSD(T) scheme is $-2.26$ Hz. In CCSD(T) the triple excitations are included fully in the fourth order in perturbation theory, and part of the 5th order terms is also included \cite{RagTruPop89}. To test the stability of this scheme, we present the results of CCSD-T calculation where further fifth-order terms are included \cite{DeeKno94}, as well as the  CCSD+T approach \cite{UrbNogBar85}, where the triple corrections are treated only at forth-order level. These values are also shown in Table ~\ref{Tab2}. While CCSD+T has the strongest effect on the calculated $W_A$ ($-2.77$ Hz), addition of the fifth order terms moderates the contribution of the triple excitations, and the difference between CCSD(T) and CCSD-T is negligible. Currently there is no possibility to evaluate the contribution of quadruple and higher excitation, but as the triple excitation contribution is already quite small, we may safely neglect the higher order ones. We take twice the difference between CCSD+T and CCSD-T ($1.5$ Hz) as the uncertainty due to incomplete treatment of correlation. We assume that the effect of replacing the Breit term by the Gaunt interaction and neglecting QED effects is not more than the contribution of the Gaunt term itself ($\sim 0.7$ Hz). The final source of uncertainty is in the numerical nature of the finite field approach, where a slight dependence on the size of the perturbation can emerge, and for small fields numerical noise might be a factor. Test calculations we carried out show that these effects are small, up to 0.5 Hz. Summing up all of the above effects we get an uncertainty estimate of $2$ Hz, or $1.5$ \%.

\begin{table}[t]
\caption{Calculated $W_\textrm{A}$ coefficient (Hz) of BaF within different correlation approaches and compared to previous predictions. The present recommended value (CCSD(T)+Gaunt) is given in bold font.}
\label{Tapr}
\begin{ruledtabular}
\begin{tabular}{lll}
$W_\textrm{A}$ (Hz)&Method&Reference\\\hline
112.32&DHF&This work\\                                                                                                                                                                                      
138.28&MP2&This work\\
150.66&CCSD&This work\\
148.40&CCSD(T)&This work\\
147.89&CCSD+T&This work\\
148.59&CCSD-T&This work\\
148.84&FSCC(0,1)-Model space I&This work\\
148.25&FSCC(0,1)-Model space II&This work\\
151.98&FSCC(1,0)&This work\\
\textbf{147.71}&\textbf{CCSD(T)+Gaunt}&\textbf{This work}\\
123.50&DFT(SVWN5)&This work\\                      
116.18&DFT(B3LYP)&This work\\
116.08&DFT(PBE)&This work\\
112.92&DFT(CAMB3LYP*)&This work\\
210-240&Semiempirical&\cite{KozLab95}\\
111&RECP-SCF&\cite{KozTitMos97}\\ 
181&RECP-SCF+EO$^a$&\cite{KozTitMos97}\\
164&Semiempirical&\cite{DeMCahMur08}\\ 
135&DHF&\cite{NayDas09}\\
160&4c-RASCI&\cite{NayDas09}\\
111&ZORA-HF&\cite{IsaBer12}\\ 
119&ZORA-DFT(B3LYP)&\cite{IsaBer12}\\
190&Scaled ZORA-HF$^b$&\cite{IsaBer12}\\
112.9&DHF&\cite{BorIliDzu13}\\
111.6&DFT(CAMB3LYP*)&\cite{BorIliDzu13}\\
146.0&DHF/DFT+CP$^c$&\cite{BorIliDzu13}\\
\end{tabular}
\end{ruledtabular}
\begin{tablenotes}
\footnotesize
\item[\emph{l}]{$^a$ RECP-SCF+EO: RECP SCF calculation with an effective operator describing valence-core correlations.}
\item[\emph{n}]{$^b$ ZORA-HF results with semiempirical scaling}
\item[\emph{c}]{$^c$ Average of DHF and DFT values, scaled by a core-polarisation parameter} 
\end{tablenotes}
\label{Tab2}
\end{table}

Table ~\ref{Tab2} ~also contains the results of the previous investigations of the $W_\textrm{A}$ parameter of BaF. The majority of these studies used approximate methods, such as DHF and DFT, or semiempirical approaches. To the best of our knowledge, this is the first investigation of this property in BaF within a relativistic coupled cluster approach, and thus direct comparison with earlier values is perhaps difficult. Our present DHF value is in excellent agreement with the RECP-SCF result of Kozlov \textit{et al.} \cite{KozTitMos97}. However, when these authors include an effective operator (EO) to account for core polarization effects, their final value overshoots the result obtained here. The DHF and DFT results of Ref. \cite{IsaBer12}  are close to the corresponding present values but the scaling scheme seems to overcompensate for the spin-polarization effects, similar to that employed in Ref. \cite{KozTitMos97}.  Our earlier DHF and DFT calculations are in good agreement with the present results, as expected, and the final value in that publication, corrected for core polarisation, is in fact very close to our CCSD(T) result, supporting the use of this scaling scheme. The DHF result of Ref. \cite{NayDas09} is larger than our value and other uncorrelated calculations \cite{KozTitMos97,IsaBer13,BorIliDzu13}, but their RASCI value is again close to our present CCSD(T) result.

\begin{figure}[t]
\includegraphics[scale=0.75,width=0.99\linewidth]{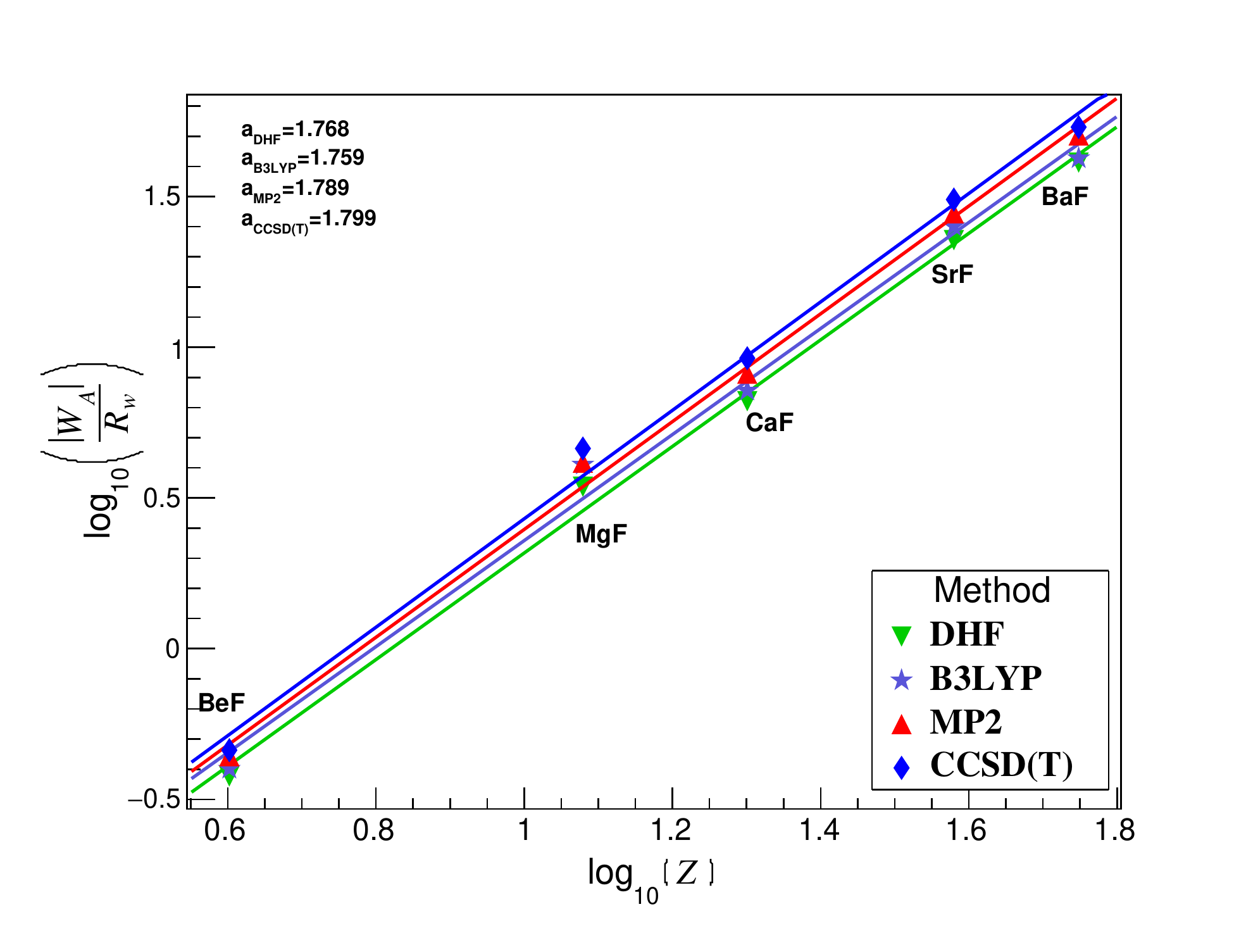}
\justifying
\begin{flushleft}
\caption{
 Scaling of $\log_{10}\left(\frac{W_\textrm{A}}{R_\textrm{W}}\right) $ with $\log_{10}(Z)$ for the selected alkaline earth fluorides.}\label{Fsca}
\end{flushleft}
\end{figure}

It is expected that the magnitude of $\frac{|W_\textrm{A}|}{R_\textrm{W}}$ of the $^2\Sigma_{1/2}$ electronic state scales as $Z^2$ \cite{FlaKhr85-1}, where the relativistic enhancement parameter $R_\textrm{W}$ ($R_\textrm{W} \ge 1$) is defined as follows \cite{FlaKhr85_2}:

\begin{eqnarray}
R_\textrm{W}   =\frac{2\gamma+1}{3}\left(  \frac{a_{B}}{2Zr_{0}A^{\frac{1}{3}}}\right)
^{2-2\gamma}\frac{4}{\left[  \Gamma(2\gamma+1)\right]  ^{2}},\label{Rw}
\end{eqnarray}

\begin{eqnarray}
\gamma & =\sqrt{1-(Z\alpha)^{2}}\label{Ga}
\end{eqnarray}
Here, $\alpha$ is the fine-structure constant, $r_{0}$ is the nucleus radius, taken here as $r_{0}=1.2\times10^{-15}$  m \cite{GinFla04}, $\Gamma(x)$ the gamma function, and $a_{B}$ is the Bohr radius. To test this dependence, we have calculated the $W_\textrm{A}$ parameters of the other alkaline earth metal fluorides. These calculations were performed with the standard Dyall's v4z basis set; all the electrons were correlated, and the energy cut-off for the virtual space was again set at 500 a.u. Table ~\ref{Tab3} ~contains the calculated $W_\textrm{A}$ parameters at the DHF, DFT, MP2, and CCSD(T) levels of theory. In Figure~\ref{Fsca}, we show $\log_{10}\left(
\frac{|W_\textrm{A}|}{R_\textrm{W}}\right) $ as a function of $\log_{10}(Z)$ for these systems. The results are fitted by a linear function: 
\begin{eqnarray}
\log_{10}\left(\frac{|W_\textrm{A}|}{R_\textrm{W}}\right) =  a\log_{10}(Z)+b.
\end{eqnarray}

For the four computational methods the  scaling factors $a$ are more or less identical ($1.77$ for DHF, $1.76$ for B3LYP, $1.79$ for MP2, and $1.80$ for CCSD(T)), in spite of very different $W_\textrm{A}$ values, implying that the trend is not sensitive to the treatment of electron correlation. Gaul \textit{et al.} report a similar finding concerning the scaling of the $P-$ and $T-$ violating parameters $W_d$ and $W_s$ in this group of molecules \cite{GauMarIsa18}.  It should be mentioned that for other sets of molecules investigated in Ref. \cite{GauMarIsa18}  (i.e. Group 4 oxides and Group 12 hydrides) this is not the case, and the Hartree-Fock and DFT scaling differ significantly. The scaling we obtain here is close (if slightly lower) to the expected $Z^2$ dependence and in good agreement with the scaling derived from the earlier DHF+DFT results \cite{BorIliDzu13} and that of Ref. \cite{IsaBer12}. In this group of molecules, no additional enhancement due to electronic structure effects is observed (unlike in group 12 fluorides, for example, where the scaling is predicted to be $2.4$ \cite{BorIliDzu13}).

\begin{table}[t]
\caption{$W_\textrm{A}$ coefficients (Hz) for alkaline earth fluorides}
\begin{ruledtabular}
\begin{tabular}{lrrrr}                                                                                                                          
Molecule&DHF&B3LYP&MP2&CCSD(T)\\\hline                                                                                                                                                                      
BeF & 0.38  &   0.40 &   0.44 &     0.46 \\ 
MgF & 3.67  &   4.34 &   4.41 &     4.91 \\
CaF & 7.74  &   8.39 &   9.55 &    10.75 \\
SrF & 37.29 &  41.50 &  45.79 &    50.87 \\
BaF & 112.27& 116.02 & 138.23 &   147.16 \\
\end{tabular}
\end{ruledtabular}
\label{Tab3}
\end{table}

\section*{Conclusions}
In this work, we presented high accuracy relativistic coupled cluster calculations for the nuclear spin dependent $P$--odd interaction constant $W_\textrm{A}$ of BaF. The effect of various computational parameters on the obtained result was explored; these include the choice of the basis set, treatment of electron correlation, number of correlated electrons, size of the virtual space, and inclusion of the Gaunt term. We find that inclusion of electron correlation raises the calculated $W_\textrm{A}$ value by about 25\%; the rest of the parameters have a much weaker effect on the results, on the order of a single percent. Furthermore, performance of various DFT functionals for this property was investigated and found lacking. We propose a final recommended value of $W_A=147.7$ Hz for BaF, obtained from the CCSD(T) calculation using the optimised basis set and corrected for the Gaunt contribution. This result, with its estimated uncertainty of 1.5\% will be useful for interpretation of future experiments on this system. We have also investigated the scaling of the $W_A$ parameter in Group 2 fluorides, and found it to be close to the expected $Z^2$ behavior.

\section*{Acknowledgement}
The authors would like to thank the Center for Information Technology of the University of Groningen for providing access to the Peregrine high performance computing cluster and for their technical support.
M.I. acknowledges the support of the Slovak Research and Development Agency and the Scientific Grant Agency, APVV-15-0105 and VEGA  1/0737/17, respectively. This research used resources of a High Performance Computing Center of the Matej Bel University in Banska Bystrica using the HPC infrastructure acquired in projects ITMS 26230120002 and 26210120002 (Slovak infrastructure for high performance computing) supported by the Research and Development Operational Programme funded by the ERDF.
PS acknowledges support from the Centre for Advanced Study at the Norwegian Academy of Science and Letters.


%

\end{document}